\documentclass{ws-procs9x6}            

\newcommand{\Eqn}[1]{&\hspace{-0.2em}#1\hspace{-0.2em}&}
\def\Vec#1{\mbox{\boldmath $#1$}}
\def\e{\mathrm{e}}

\begin{document}
\title{Inflationary universe in fluid description}

\author{Kazuharu Bamba$^*$}

\address{Division of Human Support System, Faculty of Symbiotic Systems Science, Fukushima University,\\ 
Fukushima 960-1296, Japan\\
$^*$E-mail: bamba@sss.fukushima-u.ac.jp
}

\begin{abstract}
We investigate a fluid description of inflationary cosmology. 
It is shown that the three observables of the inflationary universe: 
the spectral index of the curvature perturbations, 
the tensor-to-scalar ratio of the density perturbations, 
and the running of the spectral index, 
can be compatible with the Planck analysis. 
In addition, we reconstruct the equation of state (EoS) for a fluid 
from the spectral index of the curvature perturbations 
consistent with the Planck results. 
We explicitly demonstrate that the universe can gracefully exit from 
inflation in the reconstructed fluid models. 
Furthermore, we explore the singular inflation for a fluid model. 
\end{abstract}

\keywords{Cosmology; Particle-theory and field-theory models of the early Universe}

\bodymatter

\section{Introduction}\label{aba:sec1}

Properties of inflation\cite{Sato:1980yn,Guth:1980zm,Linde:1981mu,Albrecht:1982wi,Starobinsky:1980te} has been understood through the recent cosmological observations on the anisotropy of the cosmic microwave background (CMB) radiation by the Planck satellite\cite{Planck:2015xua,Ade:2015lrj}, the BICEP2 experiment\cite{Ade:2014xna,Ade:2015tva,Array:2015xqh}, as well as the Wilkinson Microwave anisotropy probe (WMAP)\cite{Komatsu:2010fb, Hinshaw:2012aka} (for reviews on inflation, see, for example, Refs.~\refcite{Lidsey:1995np,Lyth:1998xn,Gorbunov:2011zzc}). 

In Refs.~\refcite{Bamba:2014daa,Bamba:2014wda}, 
the three observables of the inflationary universe, namely, 
(i) the spectral index $n_\mathrm{s}$ of the curvature perturbations, 
(ii) the tensor-to-scalar ratio $r$ of the density perturbations, 
and (iii) the running $\alpha_\mathrm{s}$ of the spectral index, 
have been described with quantities in scalar field theories, 
fluid models\cite{Bamba:2012cp}, and $F(R)$ gravity theories\cite{Nojiri:2010wj,Nojiri:2006ri,Joyce:2014kja,Book-Capozziello-Faraoni,Capozziello:2011et,Koyama:2015vza,delaCruzDombriz:2012xy,Bamba:2013iga,Bamba:2014eea}. 
Inflation\cite{Barrow:1994nx,Brevik:2011mm,Wang:2013uka,Elizalde:2014ova,Brevik:2014cxa,Brevik:2014eya,Myrzakulov:2014kta,Myrzakul:2015cua,Bamba:2015sxa,Haro:2015ljc,Calzetta:1997sj,Grana:2001ms,Ackerman:2010he} and dark energy problem\cite{Nojiri:2005sr} have been considered in fluid models. 

In this paper, we review our main results in Ref.~\refcite{Bamba:2015sxa}. {}From the representation of spectral index $n_\mathrm{s}$ of the curvature perturbations in terms of the number of $e$-folds at the inflationary stage, 
which is obtained for, e.g., $R^2$ inflation (the Starobinsky inflation)\cite{Starobinsky:1980te}, we reconstruct the EoS for a fluid 
by using the procedure for scalar field theories developed in Ref.~\refcite{Chiba:2015zpa}. This representation of $n_\mathrm{s}$ can lead to its value 
compatible with the Planck data. 
It is known that in $R^2$ inflation (the Starobinsky inflation), 
the values of $n_\mathrm{s}$ and $r$ can explain the Planck analysis\cite{Planck:2015xua,Ade:2015lrj} 
(for a review of inflation in modified gravity, see, 
for instance, Ref.~\refcite{Bamba:2015uma}). 
In the reconstructed fluid models, the tensor-to-scalar ratio $r$ of the density perturbations and the running $\alpha_\mathrm{s}$ of the spectral index can 
also be consistent with the Planck results. 
Furthermore, we demonstrate that in these models, 
the universe can gracefully exit from inflation.  
We use units of $k_\mathrm{B} = c = \hbar = 1$ and express the
gravitational constant $8 \pi G_\mathrm{N}$ by
${\kappa}^2 \equiv 8\pi/{M_{\mathrm{Pl}}}^2$ 
with the Planck mass of $M_{\mathrm{Pl}} = G_\mathrm{N}^{-1/2} = 1.2 \times 
10^{19}$\,\,GeV. 

\section{Fluid description}\label{aba:sec2} 

The EoS for a general fluid can be expressed as 
$P= - \rho + f(\rho)$, 
where $\rho$ and $P$ are the energy density 
and pressure of a fluid, respectively, 
and $f(\rho)$ is a function of $\rho$. 
We assume the flat Friedmann-Lema\^{i}tre-Robertson-Walker (FLRW) metric 
$ds^2 = -dt^2 + a^2(t) d \Vec{x}^2$, where $a(t)$ is the scale factor. 
The Hubble parameter is given by $H \equiv \dot{a}/a$. 
Here, the dot denotes the derivative with respect to $t$. 
In this background space-time, 
the gravitational field equations become 
\begin{equation}
\frac{3}{\kappa^2} \left(H (N)\right)^2 = \rho(N) \,, 
\quad 
- \frac{2}{\kappa^2} H(N) \frac{dH(N)}{dN} = \rho(N) + P(N) = f(\rho) \,,
\label{eq:2.1}
\end{equation}
where the second equality of the second equation follows from the EoS. 
Here, $N$ is the number of $e$-folds during inflation, 
given by $N \equiv \ln \left(a_\mathrm{f}/a_\mathrm{i}\right)$ 
with $a_\mathrm{i}$ and $a_\mathrm{f}$ 
the values of $a(t)$ at the initial $t_\mathrm{i}$ and 
end $t_\mathrm{f}$ times of inflation, respectively. 
Moreover, the conservation law reads 
$d\rho(N)/dN + 3 \left(\rho(N) + P(N)\right) = d\rho(N)/dN + 3 f(\rho) =0$. 
Here, in deriving the first equality, we have used the EoS. 
With the second gravitational equation and the conservation law, 
we find 
\begin{equation}
\frac{2}{\kappa^2} \left(H (N)\right)^2 
\left[ \left( \frac{1}{H(N)} \frac{dH(N)}{dN}\right)^2 + \frac{1}{H(N)} \frac{d^2 H(N)}{dN^2} \right] 
= 3 \left( \frac{df(\rho)}{d\rho} \right) f(\rho) \,. 
\label{eq:2.2}
\end{equation}
It follows from Eq.~(\ref{eq:2.2}) that 
$H(N)$ and the derivatives with respect to $N$ 
can be represented by using $\rho(N)$ and $f(\rho(N))$. 
The slow-roll parameters can also be expressed with 
$\rho(N)$ and $f(\rho(N))$. 
Consequently, the three observables of the inflationary universe, 
$n_\mathrm{s}$, $r$, and $\alpha_\mathrm{s}$, 
can be described by $\rho(N)$ and $f(\rho(N))$\cite{Bamba:2014wda} 
(for the other expression of $\alpha_\mathrm{s}$, see Ref.~\refcite{Bassett:2005xm}).

\section{Reconstruction procedure of the EoS for a fluid}\label{aba:sec3} 

We study the case that the EoS for a fluid is represented as 
$P = -\rho + A \rho^{\beta}+\zeta(H)$ with $A$ and $\beta$ constants. 
Here, we take $\zeta(H) = \bar{\zeta} H^{\gamma}$ 
with $\bar{\zeta}$ and $\gamma$ constants. 
The mass dimension of $A$ and $\bar{\zeta}$ are 
$-4\left(\beta-1\right)$ and $-\left(\gamma -4 \right)$, respectively. 
Through the first equation in (\ref{eq:2.1}) for the expanding universe 
($H>0$), the EoS can be written as 
\begin{equation}  
P = - \rho + A \rho^{\beta}+\zeta(H(\rho)) 
= -\rho + A \rho^{\beta} + \bar{\zeta} \left(\frac{\kappa}{\sqrt{3}} \right)^{\gamma} \rho^{\gamma/2} \,. 
\label{eq:3.0}
\end{equation}

According to the Planck results~\cite{Planck:2015xua, Ade:2015lrj}, 
we have $n_{\mathrm{s}} = 0.968 \pm 0.006\, (68\%\,\mathrm{CL})$, 
$r< 0.11\, (95\%\,\mathrm{CL})$, 
and $\alpha_\mathrm{s} = -0.003 \pm 0.007\, (68\%\,\mathrm{CL})$. 
The constraint on $r$ is consistent with the recent 
BICEP2 and Keck Array analysis of $r< 0.09\, (95\%\,\mathrm{CL})$\cite{Array:2015xqh}. 

We consider the inflationary models in which $n_\mathrm{s}$ is described as 
$n_\mathrm{s} - 1 = -2/N$. 
This relation is satisfied in $R^2$ inflation (the Starobinsky inflation)\cite{Starobinsky:1980te}, the chaotic inflation\cite{Linde:1983gd}, 
the Higgs inflation with the non-minimal coupling\cite{Salopek:1988qh,Bezrukov:2007ep}, and the $\alpha$-attractor\cite{Kallosh:2013hoa,Kallosh:2015lwa,Kallosh:2015zsa}. 
In the case of the slow-roll inflation in a canonical scalar field theory, 
where the scalar field plays a role of an inflaton field, 
we have\cite{Chiba:2015zpa}
\begin{eqnarray} 
n_\mathrm{s} - 1 \Eqn{=} \frac{d}{d N} \left[ \ln \left( \frac{1}{V^2(N)} \frac{dV(N)}{dN} \right) \right] \,, 
\label{eq:3.1} \\
r \Eqn{=} \frac{8}{V(N)} \frac{dV(N)}{dN} \,, 
\label{eq:3.2} \\ 
\alpha_\mathrm{s} \Eqn{=} 
- \frac{d^2}{d N^2} \left[ \ln \left( \frac{1}{V^2(N)} 
\frac{dV(N)}{dN} \right) \right] \,, 
\label{eq:3.3}
\end{eqnarray}
where $V(N)$ is the inflaton potential. {}From Eq.~(\ref{eq:3.1}) and the relation $n_\mathrm{s} - 1 = -2/N$, 
we acquire $V(N) = 1/\left [\left( C_1 /N \right) + C_2 \right]$, 
where $C_1 (>0)$ and $C_2$ are constants 
and their mass dimension is four. 
For this potential, with Eq.~(\ref{eq:3.2}), we find 
$r = 8/\left\{ {N \left[1+\left(C_2/C_1\right) N \right]} \right\}$. 
In addition, by using Eq.~(\ref{eq:3.2}), 
we obtain $\alpha_\mathrm{s} = -2/N^2$. 
If $N=60$, we have $\alpha_\mathrm{s} = -5.56 \times 10^{-4}$, 
which is compatible with the Planck data. 

In the fluid description, we use the EoS for 
the inflaton potential $V$. 
We suppose that 
the slow-roll approximation that the kinetic energy is much 
smaller than the potential energy can be applied. 
In this case, we have $\rho \approx V$, and hence from 
the expression of $V$ shown above, we find 
$N \approx C_1 \rho/\left(1-C_2 \rho \right)$. 
Here, $C_2 <0$ because $N$ must be positive. 
In the FLRW space-time, 
the first equation in (\ref{eq:2.1}) with 
the slow-roll approximation, 
we obtain 
$H(N) \approx \kappa \sqrt{1/\left\{ 
3\left[ \left( C_1 /N \right) + C_2 \right] \right\}}$. 
Here, $\left( C_1 /N \right) + C_2 > 0$. 

By combining the second equation in (\ref{eq:2.1}) 
the expression of $H$ as a function of $N$, we get 
\begin{equation}  
P = -\rho - \frac{2}{\kappa^2} H(N) H'(N) 
\approx -\rho - \frac{3C_1}{N^2 \kappa^4} H^4 
\approx -\rho -\frac{1}{3C_1} \left(1-2C_2\rho +C_2^2 \rho^2 \right) \,, 
\label{eq:7}
\end{equation}
where the last approximate equality follows from the first equation in (\ref{eq:2.1}) and the relation between $N$ and $\rho$.

\section{Reconstructed EoS for a fluid}\label{aba:sec4} 

First, if $\left| C_2\rho \right| \gg 1$, it follows from Eq.~(\ref{eq:7}) 
that 
\begin{equation} 
P \approx -\rho + \frac{2C_2}{3 C_1} \rho - \frac{C_2^2}{3 C_1} \rho^2 \,. 
\label{eq:8}
\end{equation}
Moreover, we have $\left(-C_2 \right)/C_1 \approx 1/N \ll 1$, 
which follows from the relation $N \approx C_1 \rho/\left(1-C_2 \rho \right)$ 
and $\left| C_2\rho \right| \gg 1$. 
Here, $N \gg 1$ because this is the necessary condition to solve 
the so-called flatness and horizon problems. 
As a result, we obtain 
\begin{equation} 
w \equiv \frac{P}{\rho}  
\approx -1 + \frac{1}{3N} \left(-2-C_2 \rho \right) 
\approx -1\,. 
\label{eq:9}
\end{equation}
The second approximate equality in (\ref{eq:9}) can be satisfied, 
for instance, in the case that $\left| C_2\rho \right| = \mathcal{O}(10)$ and 
$N \gtrsim 60$. 
This means that the slow-roll (de Sitter) inflation can happen. 
In this case, the scale factor is represented as 
$a(t) = a_\mathrm{i} \exp \left[ H_\mathrm{inf} (t-t_\mathrm{i}) \right]$, 
where $H_\mathrm{inf} (>0)$ (= constant) 
is the Hubble parameter at the inflationary stage. 
Furthermore, when $\left(-C_2 \right)/C_1 < 1/N$ and $N \gtrsim 73$, from 
the relation $r = 8/\left\{ {N \left[1+\left(C_2/C_1\right) N \right]} 
\right\}$, we find that $r < 0.11$, namely, the constraint on the value of $r$ 
suggested by the Planck analysis can be met. 

By comparing Eqs.~(\ref{eq:3.0}) and (\ref{eq:8}), 
it is seen that these equations can be equivalent with each other 
as a linear combination of $\rho$ and $\rho^2$. 
Consequently, the following two models of 
Model (a) and Model (b) can be acquired 
\begin{eqnarray} 
&&
\mathrm{Model \,\,\, (a)}: \quad \quad \quad 
P=-\rho+ \left(\frac{2C_2}{3 C_1} \right)\rho 
- \left(\frac{3C_2^2}{C_1 \kappa^4}\right) H^4  \,, 
\label{eq:10}\\ 
&&
\mathrm{Model \,\,\, (b)}: \quad \quad \quad 
P=-\rho -\left(\frac{C_2^2}{3 C_1}\right) \rho^2 
+ \left(\frac{2C_2}{C_1 \kappa^2}\right) H^2 \,. 
\label{eq:11}
\end{eqnarray}
For Model (a), we have taken 
$A = 2C_2/\left(3 C_1\right)$, 
$\bar{\zeta} = - 3C_2^2/\left(C_1 \kappa^4\right)$, 
$\beta =1$, and $\gamma=4$, 
whereas for Model (b), we have chosen 
$A = -C_2^2/\left(3 C_1\right)$, 
$\bar{\zeta} = 2C_2/\left(C_1 \kappa^2\right)$, 
$\beta =2$, and $\gamma=2$. 

We investigate the opposite limit of $\left| C_2\rho \right| \ll 1$. {}From Eq.~(\ref{eq:7}), we find 
\begin{equation} 
P \approx -\rho 
-\frac{1}{3C_1} + \frac{2C_2}{3 C_1} \rho \,.
\label{eq:12}
\end{equation}
By using the relation $N \approx C_1 \rho/\left(1-C_2 \rho \right)$ and $\left| C_2\rho \right| \ll 1$, we get $C_1 \rho \approx N \gg 1$. 
Hence, we have $\left|C_2\right|/C_1 \ll 1$. 
With Eq.~(\ref{eq:8}), we obtain
\begin{equation} 
w = \frac{P}{\rho} 
\approx -1+ \frac{1}{3} \left( -\frac{1}{N} + 2 \frac{C_2}{C_1} \right) 
\approx -1\,,
\label{eq:13}
\end{equation}
where in deriving the first approximate equality, we have used $C_1 \rho \approx N$, and the second approximate equality follows from $N \gg 1$ and $\left|C_2\right|/C_1 \ll 1$. 
Hence, the slow-roll (de Sitter) inflation can occur. 
Moreover, if $C_2 >0$ and $C_2/C_1 \lesssim 1/N$, 
with the relation $r = 8/\left\{ {N \left[1+\left(C_2/C_1\right) N \right]} 
\right\}$, we see that when $N \gtrsim 60$, 
the tensor-to-scalar ratio $r$ of the density perturbations 
satisfies the upper limit $r < 0.11$ obtained by the Planck data. 
In addition, if $C_2 <0$ and $\left|C_2 \right|/C_1 < 1/N$, 
we find that when $N \gtrsim 73$, the upper bound $r < 0.11$ is met. 

In comparison of Eq.~(\ref{eq:12}) with Eq.~(\ref{eq:3.0}), 
the following Model (c) and Model (d) can be obtained
\begin{eqnarray} 
&&
\mathrm{Model \,\,\, (c)}: \quad \quad \quad   
P=-\rho - \left(\frac{1}{3 C_1}\right) + 
\left( \frac{2C_2}{C_1 \kappa^2} \right) H^2 \,, 
\label{eq:14}\\ 
&&
\mathrm{Model \,\,\, (d)}: \quad \quad \quad   
P=-\rho + \left(\frac{2C_2}{3 C_1} \right) \rho
- \left(\frac{1}{3 C_1}\right) \,,
\label{eq:15}
\end{eqnarray} 
For Model (c), 
$A = -1/\left(3 C_1\right)$, 
$\bar{\zeta} = 2C_2/\left(C_1 \kappa^2\right)$, 
$\beta =0$, and $\gamma=2$ have been taken, 
while for Model (d), 
$A = 2C_2/\left(3 C_1\right)$, 
$\bar{\zeta} = -1/\left(3 C_1\right)$, 
$\beta =1$, and $\gamma=0$ have been chosen. 
We note that in deriving the expressions of $\bar{\zeta}$ for Model (a), Model (b), Model (c), and Model (b), the representations of $\gamma$ for them 
have been used.

\section{Graceful exit of the universe from inflation}\label{aba:sec5} 

We demonstrate that in the reconstructed fluid models, 
the universe can gracefully exit from inflation 
by examining the instability of the de Sitter solution 
at the inflationary stage. 
We take the following perturbations of 
the Hubble parameter\cite{Bamba:2015uxa}
$H = H_\mathrm{inf} + H_\mathrm{inf} \delta(t)$ from the de Sitter solution $H_\mathrm{inf}$, where $\left| \delta(t) \right| \ll 1$, and 
therefore the second term $H_\mathrm{inf} \delta(t)$ corresponds to 
the perturbations. 
Equation (\ref{eq:2.2}) can be represented as the second differential equation with respect to $t$ as follows
\vspace{-5mm}
\begin{eqnarray}
&&
\ddot{H} -\frac{\kappa^4}{2}\left[\beta A^2 \left( \frac{3}{\kappa^2} \right)^{2\beta} H^{4\beta-1} + \left(\beta + \frac{\gamma}{2} \right) A \bar{\zeta} 
\left( \frac{3}{\kappa^2} \right)^{\beta} H^{2\beta + \gamma - 1} 
\right. \nonumber \\
&& \left. 
\hspace{15mm}
{}+ \frac{\gamma}{2} \bar{\zeta}^2 H^{2\gamma - 1} \right] = 0 \,.
\label{eq:16} 
\end{eqnarray}
As a form of $\delta(t)$ to explore the instability of the de Sitter solution, 
we define $\delta(t) \equiv \e^{\lambda t}$ with $\lambda$ a constant. 
The de Sitter solution is unstable 
when we have a positive solution of $\lambda$. 
In such a case, the graceful exit of the universe from inflation 
is realized, so that the reheating stage can follow. 
This is because the value of $\left| \delta(t) \right|$ with $\lambda > 0$ 
grows in time during inflation. 

By substituting $H = H_\mathrm{inf} + H_\mathrm{inf} \delta(t)$ 
with $\delta(t) = \e^{\lambda t}$ 
into Eq.~(\ref{eq:16}) and taking the first order of $\delta (t)$, we find
%
\begin{eqnarray}
&&
\lambda^2 - \frac{1}{2} \frac{\kappa^4}{H_\mathrm{inf}^2} \mathcal{Q} = 0 \,, 
\label{eq:17} \\
%
%
&&
\mathcal{Q} \equiv
\beta \left(4\beta - 1\right) A^2 \left( \frac{3}{\kappa^2} \right)^{2\beta} 
H_\mathrm{inf}^{4\beta} 
+ \left(\beta + \frac{\gamma}{2} \right) 
\left(2\beta + \gamma - 1\right) A \bar{\zeta} 
\left( \frac{3}{\kappa^2} \right)^{\beta} H_\mathrm{inf}^{2\beta + \gamma} 
\nonumber \\
&& 
{}+\frac{\gamma}{2} \left(2\gamma - 1\right) 
\bar{\zeta}^2 H_\mathrm{inf}^{2\gamma} \,.
\label{eq:18}
\end{eqnarray}
The solutions of Eq.~(\ref{eq:17}) read 
$\lambda = \lambda_\pm \equiv \pm \left(1/\sqrt{2}\right) 
\left(\kappa^2/H_\mathrm{inf}\right) \sqrt{\mathcal{Q}}$. 
Hence, when $\mathcal{Q} > 0$, we can have the positive solution of $\lambda = \lambda_+ > 0$. Thus, the universe can exit from inflation successfully. 

We concretely explore whether the graceful exit of the universe from inflation can occur, that is, whether $\mathcal{Q}$ can be a positive value. 
The case that the universe does not exit from inflation gracefully and 
inflation does not end corresponds to the eternal inflation. 
We substitute the values of $A$, $\bar{\zeta}$, $\beta$, and $\gamma$ 
in Models (a), (b), (c), and (d) 
into Eq.~(\ref{eq:18}), we acquire the representations of $\mathcal{Q}$ 
for these models. 
When we derive the values of $\mathcal{Q}$, 
the following points are taken into considerations. 
In all of Models (a), (b), (c), and (d), $C_1 >0$. 
In Models (a) and (b), $C_2 <0$, 
whereas in Models (c) and (d), 
$C_2$ can take both the positive value and the negative one. 

In Models (a) and (b), where $\left|C_2 \rho \right| \gg 1$, 
we obtain
\begin{eqnarray} 
&&
\mathcal{Q} = 2 \left(\frac{C_2}{C_1}\right)^2 
\left(\frac{H_\mathrm{inf}}{\kappa}\right)^4 
\left[ 6 - 45 C_2 \left(\frac{H_\mathrm{inf}}{\kappa}\right)^2 
+ 63 C_2^2 \left(\frac{H_\mathrm{inf}}{\kappa}\right)^4 \right] >0 
\nonumber\\
&&
\mathrm{for} \,\,\, 
\mathrm{Model \,\,\, (a)}\,, 
\label{eq:19} \\ 
&&
\mathcal{Q} = 6 \left(\frac{C_2}{C_1}\right)^2 
\left(\frac{H_\mathrm{inf}}{\kappa}\right)^4 
\left[ 2 - 15 C_2 \left(\frac{H_\mathrm{inf}}{\kappa}\right)^2 
+ 21 C_2^2 \left(\frac{H_\mathrm{inf}}{\kappa}\right)^4 \right] >0 
\nonumber\\
&&
\mathrm{for} \,\,\, 
\mathrm{Model \,\,\, (b)}\,.
\label{eq:20} 
\end{eqnarray}
Therefore, the condition $\mathcal{Q} >0$ is always satisfied. 
On the other hand, in Models (c) and (d), 
where $\left|C_2 \rho \right| \ll 1$, 
we find 
\begin{eqnarray} 
\hspace{-5mm}
\mathcal{Q} \Eqn{=} \left(\frac{C_2}{C_1}\right)^2 
\left(\frac{H_\mathrm{inf}}{\kappa}\right)^2
\left[ - \frac{1}{3C_2} + 12 \left(\frac{H_\mathrm{inf}}{\kappa}\right)^2 
\right] 
\quad \quad 
\mathrm{for} \,\,\, 
\mathrm{Model \,\,\, (c)}\,, 
\label{eq:21} \\ 
\hspace{-5mm}
\mathcal{Q} \Eqn{=} 2\left(\frac{C_2}{C_1}\right)^2 
\left(\frac{H_\mathrm{inf}}{\kappa}\right)^2
\left[ 6 \left(\frac{H_\mathrm{inf}}{\kappa}\right)^2 
- \frac{1}{3C_2} \right] 
\quad \quad 
\mathrm{for} \,\,\, 
\mathrm{Model \,\,\, (d)}\,.
\label{eq:22} 
\end{eqnarray}
It follows from Eqs.~(\ref{eq:21}) and (\ref{eq:22}) that 
when $C_2 <0$, we have $\mathcal{Q} >0$, 
while, for $C_2 >0$, when the following conditions are met, 
\begin{eqnarray}
\mathrm{Model \,\,\, (c)}: \quad \quad  
C_2 \Eqn{>} \frac{1}{36} \left(\frac{\kappa}{H_\mathrm{inf}}\right)^2 
\,, 
\label{eq:23} \\
\mathrm{Model \,\,\, (d)}: \quad \quad  
C_2 \Eqn{>} \frac{1}{18} \left(\frac{\kappa}{H_\mathrm{inf}}\right)^2 
\label{eq:24} 
\end{eqnarray}
we can acquire $\mathcal{Q} >0$. 
As a consequence, in the reconstructed fluid models, 
the graceful exit of the universe from inflation can be realized. 

In Models (a), (b), (c), and (d), whose EoS are given by 
Eqs.~(\ref{eq:10}), (\ref{eq:11}), (\ref{eq:14}), 
and (\ref{eq:15}), respectively, 
the three observables of the inflationary universe 
can explain the Planck data. 
The spectral index $n_\mathrm{s}$ of the curvature perturbations 
is described by $n_\mathrm{s} - 1 = -2/N$. 
It follows from this expression that if $N=60$, we have $0.967$. 
Moreover, 
the tensor-to-scalar ratio $r$ of the density perturbations 
can be smaller than the upper bound as $r < 0.11$. 
In Models (a) and (b) [$\left|C_2 \rho \right| \gg 1$] and 
Models (c) and (d) [$\left|C_2 \rho \right| \ll 1$] with $C_2 <0$, 
when $N \gtrsim 73$, we can find $r < 0.11$. 
On the other hand, in Models (c) and (d) [$\left|C_2 \rho \right| \ll 1$] 
with $C_2 >0$, for $N \gtrsim 60$, we can get $r < 0.11$. 
Furthermore, the running $\alpha_\mathrm{s}$ of the spectral index is 
expressed as $\alpha_\mathrm{s} = -2/N^2$. 
With this relation, we obtain $\alpha_\mathrm{s} = -5.56 \times 10^{-4}$. 
The values of $n_\mathrm{s}$, $r$, and $\alpha_\mathrm{s}$ shown above 
are compatible with the Planck analysis.

\section{Singular inflation for a fluid}\label{aba:sec6} 

We investigate the singular inflation\cite{Barrow:2015ora,Nojiri:2015fia,Nojiri:2015wsa,Odintsov:2015jca} for a fluid. 
This is an application of the Type IV finite-time future singularity to 
the inflationary universe. 
It is known that there are four kinds of the finite-time future singularities\cite{Nojiri:2005sx}. Their natures in modified gravity have been studied in Ref.~\refcite{Bamba:2008ut}. For the Type IV singularity, 
in the limit of $t\to t_{\mathrm{s}}$, where $t_{\mathrm{s}}$ is the time when the singularity occurs, 
the scale factor, the energy density and pressure of the universe 
do not diverge, and only the higher derivatives of $H$ becomes infinity. 
When the Type IV singularity appears, the Hubble parameter and scale factor at the inflationary stage can be represented as $H = H_\mathrm{inf} + \bar{H}\left(t_{\mathrm{s}} -t \right)^{q}$ with $q>1$ and $a = \bar{a} \exp \left\{ H_\mathrm{inf}t -\left[\bar{H}/\left(q+1\right)\right] \left(t_{\mathrm{s}} -t \right)^{q+1} \right\}$. 
Here, $\bar{H}$, $q$, and $\bar{a}$ are constants. In addition, the mass dimension of $\bar{H}$ is $q+1$. 
In this case, it is possible to show that the EoS for a fluid can approximately be a similar form to that in the reconstructed fluid models.

\section{Summary}\label{aba:sec7} 

We have executed the reconstruction of the EoS for a fluid 
from the expression of spectral index $n_\mathrm{s}$ of 
the curvature perturbations, the value of which 
is consistent with the Planck results. 
This representation is satisfied in 
$R^2$ inflation (the Starobinsky inflation). 
In the reconstructed fluid models, 
the tensor-to-scalar ratio $r$ of the density perturbations 
and the running $\alpha_\mathrm{s}$ of the spectral index can 
also explain the Planck analysis. 
Moreover, we have shown that the graceful exit from inflation can 
be realized. 
In addition, we have considered the singular inflation 
based on the Type IV singularity for a fluid model.

\section*{Acknowledgments}

The author would like to thank Professor Shin'ichi Nojiri, 
Professor Sergei D. Odintsov, and Dr. Diego S\'{a}ez-G\'{o}mez 
for our collaborations in our works\cite{Bamba:2014wda, Bamba:2015sxa} 
very much. 
This work was partially supported by the JSPS Grant-in-Aid for 
Young Scientists (B) \# 25800136 and 
the research-funds given by Fukushima University.



\begin{thebibliography}{99}

%
\bibitem{Sato:1980yn} 
  K.~Sato,
  Mon.\ Not.\ Roy.\ Astron.\ Soc.\  {\bf 195}, 467 (1981).

\bibitem{Guth:1980zm} 
  A.~H.~Guth,
  Phys.\ Rev.\ D {\bf 23}, 347 (1981).

%
\bibitem{Linde:1981mu} 
  A.~D.~Linde,
  Phys.\ Lett.\ B {\bf 108}, 389 (1982). 

\bibitem{Albrecht:1982wi} 
  A.~Albrecht and P.~J.~Steinhardt,
  Phys.\ Rev.\ Lett.\  {\bf 48}, 1220 (1982). 
%

\bibitem{Starobinsky:1980te} 
  A.~A.~Starobinsky,
  Phys.\ Lett.\ B {\bf 91}, 99 (1980).


\bibitem{Planck:2015xua} 
  P.~A.~R.~Ade {\it et al.}  [Planck Collaboration],
  arXiv:1502.01589 [astro-ph.CO].

\bibitem{Ade:2015lrj} 
  P.~A.~R.~Ade {\it et al.}  [Planck Collaboration],
  arXiv:1502.02114 [astro-ph.CO].

%
\bibitem{Ade:2014xna}
  P.~A.~R.~Ade {\it et al.}  [BICEP2 Collaboration],
  Phys.\ Rev.\ Lett.\  {\bf 112}, 241101 (2014). 

\bibitem{Ade:2015tva} 
  P.~A.~R.~Ade {\it et al.}  [BICEP2 and Planck Collaborations],
  Phys.\ Rev.\ Lett.\  {\bf 114}, 101301 (2015).

\bibitem{Array:2015xqh} 
  P.~A.~R.~Ade {\it et al.} [BICEP2 and Keck Array Collaborations],
  arXiv:1510.09217 [astro-ph.CO].

\bibitem{Komatsu:2010fb}
E.~Komatsu {\it et al.} [WMAP Collaboration],
Astrophys.\ J.\ Suppl.\ {\bf 192}, 18 (2011). 

\bibitem{Hinshaw:2012aka} 
  G.~Hinshaw {\it et al.}  [WMAP Collaboration],
  Astrophys.\ J.\ Suppl.\  {\bf 208}, 19 (2013). 

%
\bibitem{Lidsey:1995np}
  J.~E.~Lidsey, A.~R.~Liddle, E.~W.~Kolb, E.~J.~Copeland, T.~Barreiro and M.~Abney,
  Rev.\ Mod.\ Phys.\  {\bf 69}, 373 (1997). 

\bibitem{Lyth:1998xn}
  D.~H.~Lyth and A.~Riotto,
  Phys.\ Rept.\  {\bf 314}, 1 (1999). 

\bibitem{Gorbunov:2011zzc} 
D.~S.~Gorbunov and V.~A.~Rubakov, 
\textit{Introduction to the theory of the early universe: Cosmological perturbations and inflationary theory} 
(Hackensack, USA: World Scientific, 2011). 

\bibitem{Bamba:2014daa} 
  K.~Bamba, S.~Nojiri and S.~D.~Odintsov,
  Phys.\ Lett.\ B {\bf 737}, 374 (2014). 

\bibitem{Bamba:2014wda} 
  K.~Bamba, S.~Nojiri, S.~D.~Odintsov and D.~S\'{a}ez-G\'{o}mez, 
  Phys.\ Rev.\ D {\bf 90}, 124061 (2014). 

\bibitem{Bamba:2012cp} 
  K.~Bamba, S.~Capozziello, S.~Nojiri and S.~D.~Odintsov,
  Astrophys.\ Space Sci.\  {\bf 342}, 155 (2012)
  [arXiv:1205.3421 [gr-qc]].

%
\bibitem{Nojiri:2010wj}
S.~Nojiri and S.~D.~Odintsov,
Phys.\ Rept.\ {\bf 505}, 59 (2011).

\bibitem{Nojiri:2006ri}
S.~Nojiri and S.~D.~Odintsov,
eConf C {\bf 0602061} (2006) 06  
[Int.\ J.\ Geom.\ Meth.\ Mod.\ Phys.\ {\bf 4}, 115 (2007)].

\bibitem{Joyce:2014kja} 
  A.~Joyce, B.~Jain, J.~Khoury and M.~Trodden,
  Phys.\ Rept.\  {\bf 568}, 1 (2015).

\bibitem{Book-Capozziello-Faraoni}
S.~Capozziello and V.~Faraoni,
\textit{Beyond Einstein Gravity}
(Springer, Dordrecht, 2010). 

\bibitem{Capozziello:2011et}
S.~Capozziello and M.~De Laurentis,
Phys.\ Rept.\ {\bf 509}, 167 (2011). 

\bibitem{Koyama:2015vza} 
  K.~Koyama,
  arXiv:1504.04623 [astro-ph.CO].

\bibitem{delaCruzDombriz:2012xy}
  A.~de la Cruz-Dombriz and D.~S\'{a}ez-G\'{o}mez,
  Entropy {\bf 14}, 1717 (2012).

\bibitem{Bamba:2013iga} 
  K.~Bamba, S.~Nojiri and S.~D.~Odintsov,
  arXiv:1302.4831 [gr-qc].

\bibitem{Bamba:2014eea}
   K.~Bamba and S.~D.~Odintsov,
   arXiv:1402.7114 [hep-th]. 


\bibitem{Barrow:1994nx} 
  J.~D.~Barrow and J.~P.~Mimoso,
  Phys.\ Rev.\ D {\bf 50}, 3746 (1994).

\bibitem{Brevik:2011mm} 
  I.~Brevik, E.~Elizalde, S.~Nojiri and S.~D.~Odintsov,
  Phys.\ Rev.\ D {\bf 84}, 103508 (2011). 

\bibitem{Wang:2013uka} 
  J.~Wang and X.~Meng,
  Mod.\ Phys.\ Lett.\ A {\bf 29}, 1450009 (2014). 

\bibitem{Elizalde:2014ova} 
  E.~Elizalde, V.~V.~Obukhov and A.~V.~Timoshkin,
  Mod.\ Phys.\ Lett.\ A {\bf 29}, 1450132 (2014). 

\bibitem{Brevik:2014cxa} 
  I.~Brevik and {\O}.~Gr{\o}n,
  arXiv:1409.8561 [gr-qc].

\bibitem{Brevik:2014eya} 
  I.~Brevik, V.~V.~Obukhov and A.~V.~Timoshkin,
  Astrophys.\ Space Sci.\  {\bf 355}, 399 (2015). 

\bibitem{Myrzakulov:2014kta} 
  R.~Myrzakulov and L.~Sebastiani,
  Astrophys.\ Space Sci.\  {\bf 356}, 205 (2015). 

\bibitem{Myrzakul:2015cua} 
  S.~Myrzakul, R.~Myrzakulov and L.~Sebastiani,
  Astrophys.\ Space Sci.\  {\bf 357}, 168 (2015). 

\bibitem{Bamba:2015sxa} 
  K.~Bamba and S.~D.~Odintsov,
  arXiv:1508.05451 [gr-qc].

\bibitem{Haro:2015ljc} 
  J.~Haro and S.~Pan,
  arXiv:1512.03033 [gr-qc].

\bibitem{Nojiri:2005sr} 
  S.~Nojiri and S.~D.~Odintsov,
  Phys.\ Rev.\ D {\bf 72}, 023003 (2005). 

\bibitem{Calzetta:1997sj} 
  E.~Calzetta and M.~Grana,
  astro-ph/9705069.

\bibitem{Grana:2001ms} 
  M.~Grana and E.~Calzetta,
  Phys.\ Rev.\ D {\bf 65}, 063522 (2002). 

\bibitem{Ackerman:2010he} 
  L.~Ackerman, W.~Fischler, S.~Kundu and N.~Sivanandam,
  JCAP {\bf 1105}, 024 (2011). 


\bibitem{Chiba:2015zpa} 
  T.~Chiba,
  PTEP {\bf 2015}, 073E02 (2015). 

\bibitem{Bamba:2015uma} 
  K.~Bamba and S.~D.~Odintsov,
  Symmetry {\bf 7}, 220 (2015). 


\bibitem{Bassett:2005xm}
  B.~A.~Bassett, S.~Tsujikawa and D.~Wands,
  Rev.\ Mod.\ Phys.\  {\bf 78}, 537 (2006). 


\bibitem{Linde:1983gd} 
  A.~D.~Linde,
  Phys.\ Lett.\ B {\bf 129}, 177 (1983).

%
\bibitem{Salopek:1988qh} 
  D.~S.~Salopek, J.~R.~Bond and J.~M.~Bardeen,
  Phys.\ Rev.\ D {\bf 40}, 1753 (1989). 

\bibitem{Bezrukov:2007ep} 
  F.~L.~Bezrukov and M.~Shaposhnikov,
  Phys.\ Lett.\ B {\bf 659}, 703 (2008). 
%

%
\bibitem{Kallosh:2013hoa} 
  R.~Kallosh and A.~Linde,
  JCAP {\bf 1307}, 002 (2013). 

\bibitem{Kallosh:2015lwa} 
  R.~Kallosh and A.~Linde,
  Phys.\ Rev.\ D {\bf 91}, 083528 (2015). 

\bibitem{Kallosh:2015zsa} 
  R.~Kallosh and A.~Linde,
  Comptes Rendus Physique {\bf 16}, 914 (2015)

\bibitem{Bamba:2015uxa} 
  K.~Bamba, S.~D.~Odintsov and P.~V.~Tretyakov,
  Eur.\ Phys.\ J.\ C {\bf 75}, 344 (2015). 

\bibitem{Barrow:2015ora} 
  J.~D.~Barrow and A.~A.~H.~Graham,
  Phys.\ Rev.\ D {\bf 91}, 083513 (2015). 

\bibitem{Nojiri:2015fia} 
  S.~Nojiri, S.~D.~Odintsov, V.~K.~Oikonomou and E.~N.~Saridakis,
  JCAP {\bf 1509}, 044 (2015).

\bibitem{Nojiri:2015wsa} 
  S.~Nojiri, S.~D.~Odintsov and V.~K.~Oikonomou,
  Phys.\ Lett.\ B {\bf 747}, 310 (2015). 

\bibitem{Odintsov:2015jca} 
  S.~D.~Odintsov and V.~K.~Oikonomou,
  Phys.\ Rev.\ D {\bf 92}, 024058 (2015).

\bibitem{Nojiri:2005sx} 
  S.~Nojiri, S.~D.~Odintsov and S.~Tsujikawa,
  Phys.\ Rev.\ D {\bf 71}, 063004 (2005).

\bibitem{Bamba:2008ut} 
  K.~Bamba, S.~Nojiri and S.~D.~Odintsov,
  JCAP {\bf 0810}, 045 (2008).


\end{thebibliography}
\end{document}